\documentclass[a4paper]{article}
\usepackage{INTERSPEECH2022}
\usepackage{multirow}
\usepackage{booktabs}
\usepackage{makecell}
\usepackage{amsmath}
\usepackage{color}

\title{Reliable Visualization for Deep Speaker Recognition}
\name{Pengqi Li$^1$, Lantian Li$^2$, Askar Hamdulla$^1$, Dong Wang$^{2}$
\thanks{
This work was supported by the National Natural Science
Foundation of China under Grant No.62171250,
and also the Tsinghua-SPD Bank Joint Research Center for Digital Finance Technologies.
Dong Wang is the corresponding author.}}
\address{
  $^1$Information Science and Engineering Institute, Xinjiang University \\
  $^2$Center for Speech and Language Technologies, Tsinghua University}
\email{\{lipq,lilt\}@cslt.org, askar@xju.edu.cn, wangdong99@mails.tsinghua.edu.cn}

\begin{document}

\maketitle

\begin{abstract}
In spite of the impressive success of convolutional neural networks (CNNs) in speaker recognition,
our understanding to CNNs' internal functions is still limited. A major obstacle is 
that some popular visualization tools are difficult to apply, for example those producing saliency maps.
The reason is that speaker information does not show clear spatial patterns in the temporal-frequency 
space, which makes it hard to interpret the visualization results, and hence hard to confirm the reliability of 
a visualization tool.
  
In this paper, we conduct an extensive analysis on three popular visualization methods based on CAM: 
Grad-CAM, Score-CAM and Layer-CAM, to investigate their reliability for speaker recognition tasks.
Experiments conducted on a state-of-the-art ResNet34SE model show that the Layer-CAM algorithm
can produce reliable visualization, and thus can be used as a promising tool to explain CNN-based speaker models.
The source code and examples are available in our project page: \emph{http://project.cslt.org/}.
\end{abstract}

\section{Introduction}

Deep convolutional neural networks (CNNs) have attained remarkable success in 
computer vision~\cite{lee2009convolutional,murphy2016overview,gu2018recent}. Besides the 
unprecedented performance on a broad range of tasks, a special reason is that there are 
multiple visualization tools that can be used to explain the decisions of the 
model~\cite{simonyan2014very,ribeiro2016should,zhou2016learning}. 
Some of the most representative visualization tools include guided backpropagation~\cite{springenberg2014striving},
deconvolution~\cite{zeiler2014visualizing}, CAM and its variants~\cite{zhou2016learning,chattopadhay2018grad,wang2020score,jiang2021layercam}.
These tools can produce saliency maps that identify the important regions in an image that lead to the model's decision.
Importantly, humans can easily interpret a saliency map of an image, and hence judge the quality of a visualization tool.

Recently, CNN models have been widely adopted in speaker recognition and achieved fairly 
good performance~\cite{bai2021speaker}.
However, how the models obtain such performance is hard to explain. 
A major obstacle is that the established visualization tools cannot be used directly.
Basically, this is because humans cannot `see' speech, which makes interpretation and 
quality judgement for saliency maps on speech signals quite difficult~\cite{krug2018introspection}.
Recently, some researchers have recognized the difficulty, and provide some 
solutions~\cite{trinh2022identifying,krug2018introspection,trinh2020large}.
Nearly all the research focused on visualizing phone classes.

So far very few research in literature reports visualization for speaker recognition. 
This is not surprising as speaker traits spread among nearly all frequency bands and 
temporal segments, making the question `where is important' difficult to answer. In contrast, phone 
information is much more localized, at least in the temporal axis.  
Among the few exceptions, \cite{zhou2021resnext} employed Grad-CAM~\cite{selvaraju2017grad} 
to compare the behavior of ResNet and Res2Net under noisy corruption. They found that 
the saliency map produced by Grad-CAM is more stable with Res2Net compared with ResNet, thus explaining the 
advantage of Res2Net. In~\cite{himawan2019voice}, the authors used Grad-CAM to analyze 
genuine speech and spoof speech, and found that CNN models look into high-frequency 
components to identify spoof speech. All the mentioned studies employed the visualization tools directly
by assuming that they are correct. Unfortunately,
\emph{so far we have no idea if any of the visualization tools are reliable when applied to 
speaker recognition, which makes the conclusions obtained from visualization not fully convincing.} 

In this paper, we focus on the popular CAM-based algorithms~\cite{zhou2016learning}, and try to answer the question
\emph{if these algorithms, or any of them, can be reliably applied to speaker recognition tasks}.
Three CAM algorithms will be investigated: Grad-CAM++~\cite{chattopadhay2018grad}, 
Score-CAM~\cite{wang2020score} and Layer-CAM~\cite{jiang2021layercam}.
The main idea of these algorithms is to generate a saliency map by 
combing the activation maps (channels) of a convolutional layer. 

Our investigation starts from a deletion and insertion experiment, as suggested in~\cite{petsiuk2018rise}. 
In the deletion process, the most relevant regions (MoRRs) in Mel spectrograms are gradually masked by setting the values to $0$
according to a saliency map. 
In the insertion process, the MoRRs are gradually unmasked (exposed), starting 
from a totally-masked Mel spectrogram. For each insertion and deletion config, we examine the accuracy of a 
speaker recognition model. This results in a deletion curve and an insertion curve, by which we can tell if 
a saliency map really identifies the salient regions, and compare quality of different saliency maps, hence different 
visualization algorithms. 

We conducted experiments with a ResNet34SE x-vector model. 
The results show that all the three CAM algorithms outperform random masking and time-aligned masking, demonstrating that they 
are effective. Among the three CAMs, Layer-CAM shows superiority in the deletion/insertion test, and produces more localized 
patterns. However, no clear temporal-frequency (T-F) patterns are detected.

Further more, we conducted the deletion/insertion experiment on multi-speaker speech. This time, Layer-CAM demonstrates surprisingly
good performance in distinguishing target speakers and interfering speakers, and the other two CAMs largely fail. This clearly 
shows that only Layer-CAM is a valid visualization tool for speaker recognition. 


\section{Related work}
\label{sec:work}

Understanding the behavior of deep speaker recognition systems  by visualization is a common practice.
A popular approach is to visualize the distributions of 
frame-level or utterance-level representations~\cite{li2017deep,shon2018frame,jia2018transfer,xu2018generative,bhattacharya2019deep,shi2020h},
via manifold learning algorithms~\cite{weinberger2006unsupervised} such as PCA and t-SNE~\cite{van2008visualizing}.
Some researchers conduct visualization based on scores of trials. 
For instance, \cite{kinnunen2021visualizing} visualizes the relation of different 
systems by \emph{multidimensional scaling} (MDS), where the distance between 
a pair of systems is derived from the scores produced by each of them. 
Recently, the authors proposed a novel config-performance (C-P) map tool that
visualizes the performance of an ASV system in a 2-dimensional map~\cite{li2022cp}.
All these visualization methods can (partly) show the behavior of a system, but provide 
little \emph{understanding} for the behavior. CAM, or other saliency-map algorithms,
is supposed to offer better explanations for a CNN model, hence better understanding.

The saliency-map algorithms can be categorized into three classes:
gradient-based approach~\cite{simonyan2014very,springenberg2014striving,zeiler2014visualizing,sundararajan2017axiomatic,adebayo2018local}, 
perturbation-based approach~\cite{zeiler2014visualizing,petsiuk2018rise,fong2017interpretable,chang2018explaining} 
and CAM-based approach~\cite{zhou2016learning,chattopadhay2018grad,wang2020score,jiang2021layercam,selvaraju2017grad}.
The gradient-based approach is subject to low quality and noisy interference~\cite{omeiza2019smooth},
and the perturbation-based approach usually needs additional regularizations \cite{fong2017interpretable} and is time-costing.
In comparison, the CAM-based approach often creates more clear and localized saliency maps, thus being adopted widely by the 
computer vision community. 

Considering the high quality and localization capability,
this paper focuses on CAMs, with the goal of identifying a CAM algorithm that can 
be reliably used to visualize speaker recognition models.
To the best of our knowledge, this is the first work towards this direction.





\section{Methodology}
\label{sec:method}

A class activation map (CAM) is a saliency map that shows the important regions used by the CNN to identify a particular class.
In this section, we will firstly revisit three CAM algorithms that we will experiment with, 
and then present the normalization process designed to attain suitable T-F masks.

\subsection{Revisit CAMs}

\subsubsection{Grad-CAM and Grad-CAM++}

We start from Grad-CAM~\cite{selvaraju2017grad}.
Let $f$ denote the speaker classifier instantiated by a CNN, and $\theta$ represents its parameters.
For a given input $x$ from class $c$, the prediction score (posterior probability) for the target class 
can be computed by a forward pass:

\begin{equation}
\label{eq:pred}
y^c = f_c(x; \theta).
\end{equation}
\noindent Then for the $k$-th activation map (i.e., the $k$-th channel) $A^k$ of a convolutional layer,
the gradient of $y^c$ with respect to $A^k_{ij}$ is computed and the values at all the locations are 
averaged to obtain the weight for $A^k$ with respect to class $c$:

\begin{equation}
\label{eq:weight}
w_{k}^{c} = \frac{1}{Z} \sum_{i} \sum_{j} \frac{\partial y^{c}}{\partial A_{i j}^{k}},
\end{equation}
\noindent where $Z$ is a constant corresponding to the number of points in the map.
Grad-CAM then produces the saliency map for class $c$ by linearly combining $A^k$ with weight $w_k^c$
and followed by $relu(\cdot)$:

\begin{equation}
\label{eq:smap}
S^{c} = relu(\sum_{k}w_{k}^{c}\cdot A^{k})).
\end{equation}


Grad-CAM++~\cite{chattopadhay2018grad} is a derived version of Grad-CAM, where the weight $w_k^c$ for $A^k$ is computed by:

\begin{equation} \label{GradCAM++_function}
w_{k}^{c}= \frac{1}{Z} \sum_{i}\sum_{j}\alpha_{ij}^{kc}\cdot relu(\frac{\partial y^{c}}{\partial A_{i j}^{k}}).
\end{equation}

Specifically, Grad-CAM++ focuses on positive gradients only, and weights $\frac{\partial y^{c}}{\partial A_{i j}^{k}}$
by $\alpha_{ij}^{kc}$. This change helps identify multiple occurrences of the same class.

\subsubsection{Score-CAM}

Gradients of a deep neural network can be noisy and vanished.
Score-CAM is a gradient-free algorithm~\cite{wang2020score}, which computes the weights $w_k^c$ for activation map $A^k$ 
by forward activation rather than backward gradient. 
Specifically, it firstly forwards $x$ through the CNN to generate activation map $\{A^k\}$, and then use $A^k$ to mask $x$:

\begin{equation}
\label{eq:score}
\hat{x}_k =  x \circ \{\text{Norm}(\text{Upsampling}(A^k)) \}
\end{equation}
\noindent where Upsampling($\cdot$) stretches $A^k$ to meet the size of $x$, and Norm($\cdot$) performs a min-max normalization.
$\hat{x}_k$ is then passed through the CNN again, and the generated posterior probability $f_c(\hat{x}_k)$ is used as the weight $w_k^c$.

\subsubsection{Layer-CAM}

Layer-CAM~\cite{jiang2021layercam} is gradient-based, and reweights $A^k$ in a point-wised way.
The weight for activation map $A^k$ for class $c$ at location ($i,j$) is defined as the gradient at that location:

\begin{equation}
\label{eq:layer-w}
w_{ij}^{kc} = relu(\frac{\partial y^{c}}{\partial A_{i j}^{k}}).
\end{equation}

\noindent Note that only positive gradients are considered, as in Grad-CAM++. The saliency map is produced as follows:

\begin{equation}
\label{eq:layer}
S_{ij}^c = relu \{ \sum_k w_{ij}^{kc} \cdot {A_{i j}^{k}} \}.
\end{equation}

\noindent It was demonstrated that the point-wised weighting is beneficial to produce more fine-grained and localized saliency maps~\cite{jiang2021layercam}.

\subsection{Saliency normalization}

The values in the saliency map $S^c$ may vary in a large range, and the range 
could be very different with different CAMs. To make the maps comparable in visualization, 
and to use them for speaker localization (Ref. Section~\ref{sec:exp}), we firstly re-arrange the saliency 
values into the interval [0,1] by min-max normalization:

\begin{equation}
\label{eq:norm}
\hat{S}^{c} = \frac{S^{c}-\min S^{c}}{\max S^{c}-\min S^{c}}.
\end{equation}
\noindent Furthermore, a scale function is applied to redistribute the saliency values.
According to~\cite{jiang2021layercam}, we choose $tanh$ as the scale function:

\begin{equation} \label{scale}
\hat{S}^{c}_n = tanh(\frac{\gamma * \hat{S}^{c}}{\max \hat{S}^{c}}),
\end{equation}
\noindent where $\gamma$ is the scale coefficient which was set to be $5$ in our experiments. 

\section{Experiments}
\label{sec:exp}

In this section, we quantitatively evaluate the reliability of three CAM algorithms: Grad-CAM++, 
Score-CAM and Layer-CAM, using a well-trained deep speaker model.

\subsection{Speaker model}

The development set of VoxCeleb2~\cite{nagrani2020voxceleb} was used to train the speaker model,
which contains 5,994 speakers in total. No data augmentation was used.
The structure of the model is ResNet34 with squeeze-and-excitation (SE) layers~\cite{hu2018squeeze},
shown in Table~\ref{tab:model}. 
The model was trained by the Adam optimizer, following the voxceleb/v2 recipe of the Sunine toolkit~\footnote{https://gitlab.com/csltstu/sunine}.

\begin{table}[htb!]
\centering
\caption{The topology of ResNet34SE model.}
\label{tab:model}
\scalebox{0.9}{
\begin{tabular}{lcc}
\toprule
\toprule
\textbf{Layer}     & \textbf{Module}                           & \textbf{Output}          \\
\midrule
Input              & --                                        & 80$\times$200$\times$1   \\
Conv2D             & 3$\times$3$\times$32, \text{Stride} 1     & 80$\times$200$\times$32  \\
\midrule
ResNetBlock1       & $\begin{bmatrix}\text{3}\times\text{3}\times\text{32} \\ \text{3}\times\text{3}\times\text{32}  \\ \text{SE Layer}\end{bmatrix}$$\times$3, \text{Stride} 1    & 80$\times$200$\times$32  \\
\midrule
ResNetBlock2       & $\begin{bmatrix}\text{3}\times\text{3}\times\text{64}  \\ \text{3}\times\text{3}\times\text{64} \\ \text{SE Layer}\end{bmatrix}$$\times$4, \text{Stride} 2    & 40$\times$100$\times$64  \\
\midrule
ResNetBlock3       & $\begin{bmatrix}\text{3}\times\text{3}\times\text{128} \\ \text{3}\times\text{3}\times\text{128} \\ \text{SE Layer}\end{bmatrix}$$\times$6, \text{Stride} 2    & 20$\times$50$\times$128  \\
\midrule
ResNetBlock4       & $\begin{bmatrix}\text{3}\times\text{3}\times\text{256} \\ \text{3}\times\text{3}\times\text{256} \\ \text{SE Layer}\end{bmatrix}$$\times$3, \text{Stride} 2    & 10$\times$25$\times$256  \\
\midrule
Pooling            & TSP~\cite{snyder2018x}               & 20$\times$256        \\
Flatten            & --                                   & 5120                 \\
\midrule
Dense              & --                                   & 256                  \\
Dense              & AM-Softmax~\cite{wang2018additive}   & 5994                 \\
\bottomrule
\bottomrule
\end{tabular}}
\end{table}

\subsection{Single-speaker experiment}

In this section, we probe the behavior of the three CAM algorithms using
single-speaker utterances, i.e., only a single target speaker exists in an utterance. Our purpose is to examine if 
the CAM algorithms can detect salient components in Mel spectrograms.
We randomly choose 2,000 utterances of 200 speakers (10 utterances per speaker) from the training data to perform the evaluation.
The saliency maps are generated from ResNetBlock4 (S4), by Grad-CAM++, Score-CAM and Layer-CAM respectively.

\begin{figure}[!htbp]
\centering
\includegraphics[width=1\linewidth]{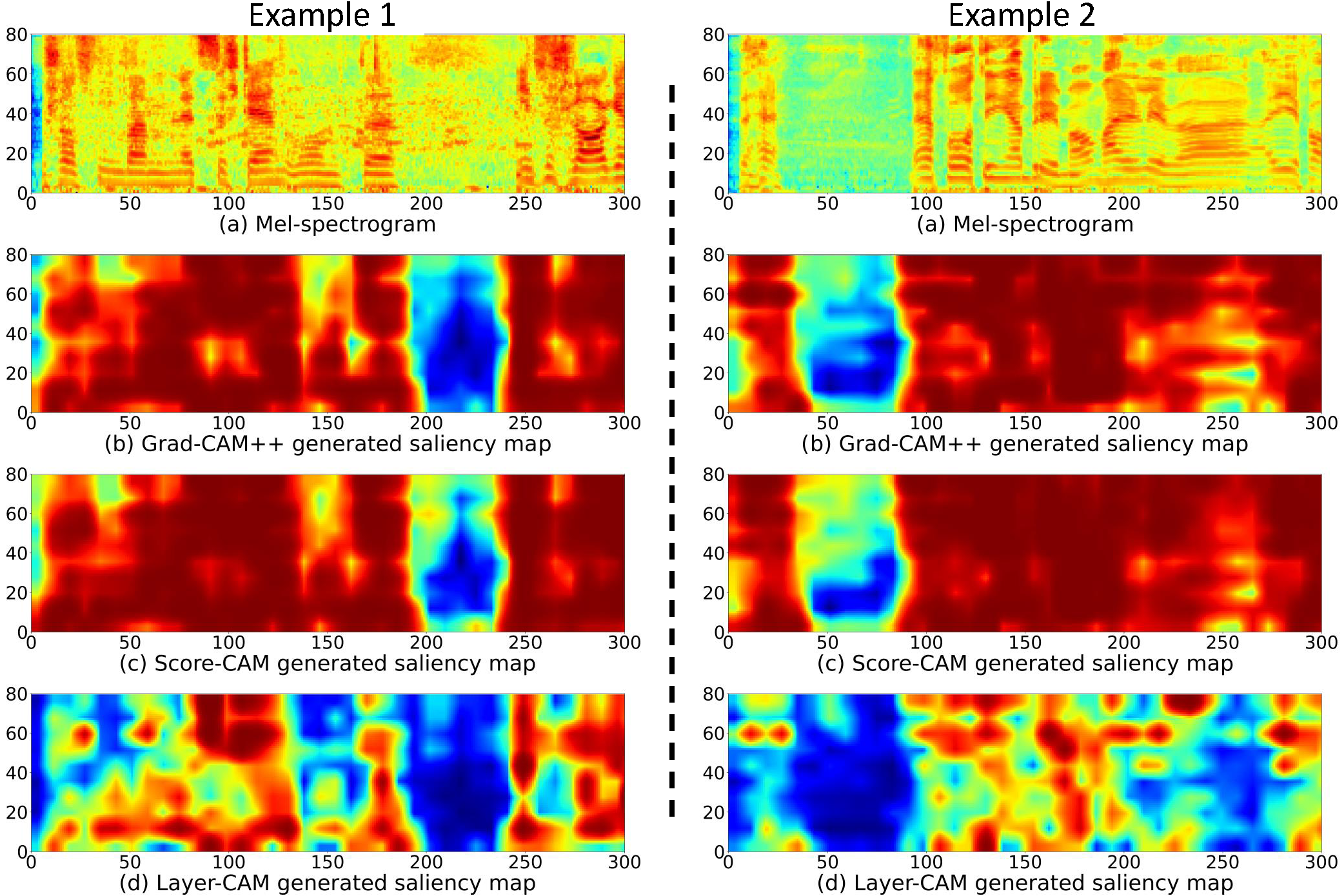}
\caption{The saliency maps of two speaker utterances generated by GradCAM++, Score-CAM and Layer-CAM. The deeper the color, the more important the region.
}
\label{fig:map}
\end{figure}

First of all, we qualitatively compare the saliency maps produced by the three CAM-based methods.
Two examples are shown in Figure~\ref{fig:map}. 
It can be observed that all the saliency maps clearly separate speech and non-speech segments,
demonstrating the basic capacity of the CAMs. 
Another observation is that Grad-CAM++ and Score-CAM tend to regard all the speech segments being important,
while Layer-CAM produces more selective and localized patterns. Nevertheless, no clear T-F patterns are found in any of the saliency maps.
More examples are provided in our project page\footnote{http://project.cslt.org/}.

\begin{table*}[htb!]
\centering
\caption{Top-1 Acc (\%) on target-speaker localization and recognition task with three CAMs.
A represents target speaker; B and C represent non-target speakers. G-CAM: Grad-CAM++; S-CAM: Score-CAM; L-CAM: Layer-CAM.
S1$\sim$S4 represents ResNetBlock1$\sim$ResNetBlock4.`+' denotes element-wised average of multiple saliency maps.}
\label{tab:local}
\scalebox{0.86}{
\begin{tabular}{lcccccccccccc}
\toprule
\textbf{Cases}         & \multicolumn{3}{c}{\textbf{A-B}}       & \multicolumn{3}{c}{\textbf{A-B-A}}     & \multicolumn{3}{c}{\textbf{B-A-B}}        & \multicolumn{3}{c}{\textbf{A-B-C}}       \\
\cmidrule(r){1-1}      \cmidrule(r){2-4}                        \cmidrule(r){5-7}                          \cmidrule(r){8-10}                            \cmidrule(r){11-13}
\textbf{Original}      & \multicolumn{3}{c}{49.15\%}            & \multicolumn{3}{c}{83.55\%}              & \multicolumn{3}{c}{15.35\%}                 & \multicolumn{3}{c}{30.30\%}              \\
\cmidrule(r){1-1}      \cmidrule(r){2-4}                        \cmidrule(r){5-7}                          \cmidrule(r){8-10}                            \cmidrule(r){11-13}
\textbf{Settings}      & G-CAM    & S-CAM    & L-CAM            & G-CAM     & S-CAM     & L-CAM            & G-CAM     & S-CAM     & L-CAM               & G-CAM     & S-CAM     & L-CAM            \\
\cmidrule(r){1-1}      \cmidrule(r){2-4}                        \cmidrule(r){5-7}                          \cmidrule(r){8-10}                            \cmidrule(r){11-13}
\textbf{S1}            & 43.00\%  & 34.00\%  & 6.75\%           & 75.55\%   & 62.50\%   & 8.90\%           & 15.15\%   & 12.10\%   & 4.15\%              & 22.40\%   & 17.45\%   & 3.90\%           \\
\textbf{S2}            & 46.60\%  & 46.60\%  & 61.85\%          & 79.90\%   & 79.25\%   & 85.00\%          & 15.70\%   & 16.00\%   & 35.20\%             & 26.85\%   & 26.85\%   & 45.15\%          \\
\textbf{S3}            & 48.45\%  & 48.60\%  & 49.40\%          & 82.65\%   & 82.35\%   & 80.15\%          & 15.75\%   & 16.00\%   & 20.20\%             & 29.20\%   & 29.55\%   & 31.05\%          \\
\textbf{S4}            & 49.20\%  & 48.25\%  & 53.20\%          & 82.10\%   & 82.65\%   & 82.90\%          & 17.20\%   & 16.15\%   & 24.05\%             & 30.10\%   & 29.20\%   & 34.65\%          \\
\textbf{S4+S3}         & 48.65\%  & 48.15\%  & 51.15\%          & 82.50\%   & 82.25\%   & 82.60\%          & 16.50\%   & 16.15\%   & 21.90\%             & 29.40\%   & 29.30\%   & 33.55\%          \\
\textbf{S4+S3+S2}      & 48.55\%  & 48.40\%  & 59.85\%          & 82.20\%   & 82.00\%   & 87.30\%          & 16.10\%   & 16.20\%   & 28.65\%             & 29.65\%   & 29.15\%   & 42.75\%          \\
\textbf{S4+S3+S2+S1}   & 47.70\%  & 47.50\%  & \textbf{71.55\%} & 81.50\%   & 80.65\%   & \textbf{92.20\%} & 16.10\%   & 16.10\%   & \textbf{44.60\%}    & 27.95\%   & 27.45\%   & \textbf{58.90\%} \\
\bottomrule
\end{tabular}}
\end{table*}





\begin{figure}[htb!]
  \centering
  \includegraphics[width=1\linewidth]{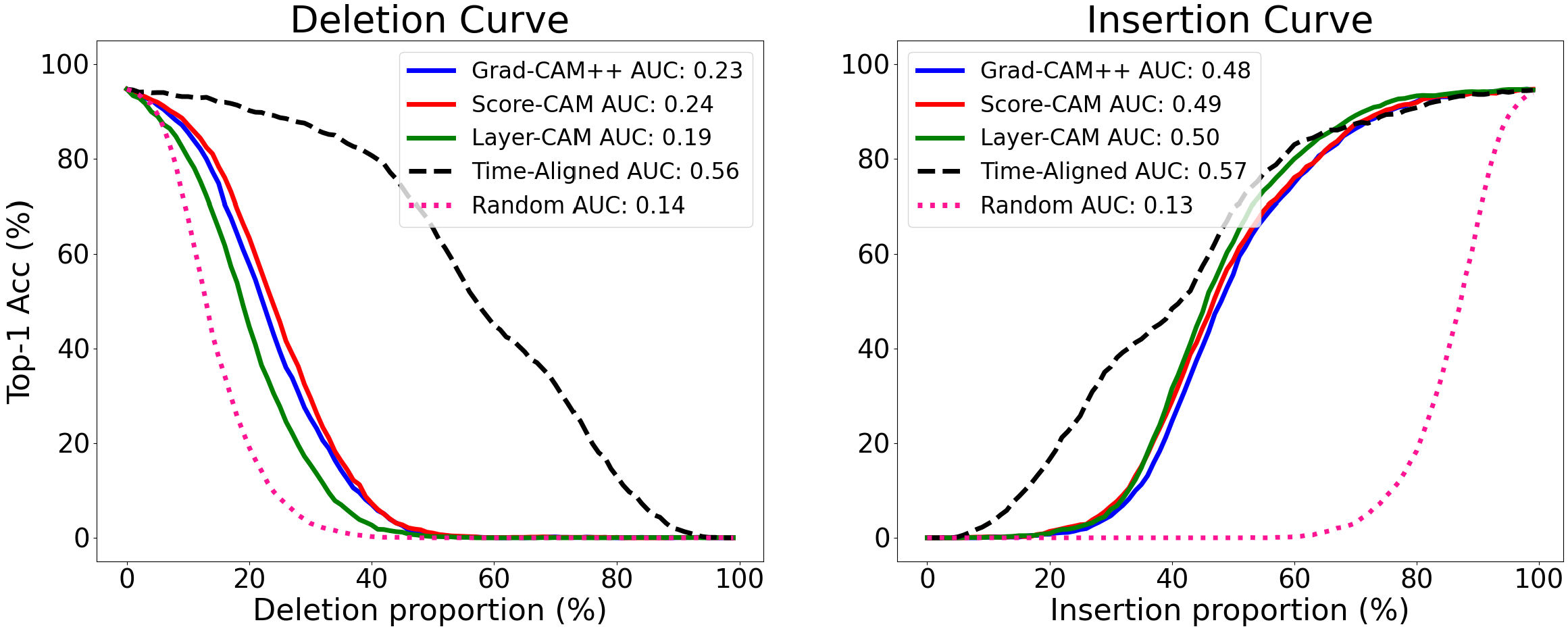}
  \caption{Deletion and insertion curves of three CAM algorithms with single-speaker speech.}
  \label{fig:auc}
\end{figure}

We further conduct the deletion and insertion experiment suggested in~\cite{petsiuk2018rise}.
In the deletion experiment, we monitor the Top-1 accuracy of the CNN model as more and more important regions of the input are masked,
and in the insertion experiment, we monitor the Top-1 accuracy as more and more important regions are unmasked.
The Area Under the Curve (AUC) value is used to measure the quality of the saliency maps. 
More the saliency map is accurate, more the AUC lower in the deletion curve and higher in the insertion curve. 
Since we have known that all the CAM algorithms are capable of detecting non-speech segment, 
we focus on the speech segment in this experiment. For that purpose, voice activity detection (VAD) has been firstly employed to remove non-speech segments. 
Moreover, we show the performance with random (un)masking and left-to-right time-aligned (un)masking as reference (un)masking methods. 

Figure~\ref{fig:auc} shows the results. It can be observed that the three CAMs are comparable in this deletion/insertion test, though Layer-CAM is slightly better. 
The comparison with random masking and time-aligned masking is also interesting: it shows that 
the three CAM algorithms indeed find salient regions. 
For example, in the insertion experiment, the curves of CAM algorithms clearly are much higher than that of the random masking, 
indicating that the regions exposed earlier by CAMs are indeed more important than random regions. 
And in the deletion experiment, the curves of CAM algorithms are much lower than that of the time-aligned masking, showing 
that the regions deleted in the early stage by CAMs are more important than real speech with the same amount of T-F bins. 
Note that the quick accuracy increase with time-aligned masking in the insertion curve is understandable, as 
it inserts entire frames which are valuable for speaker recognition when the utterance is short. 
Similarly, the quick accuracy drop with random masking in the deletion curve is not surprising, 
as random noise is very harmful for speaker recognition.

\subsection{Multi-speaker experiment}

\begin{figure}[!htbp]
\centering
\includegraphics[width=1\linewidth]{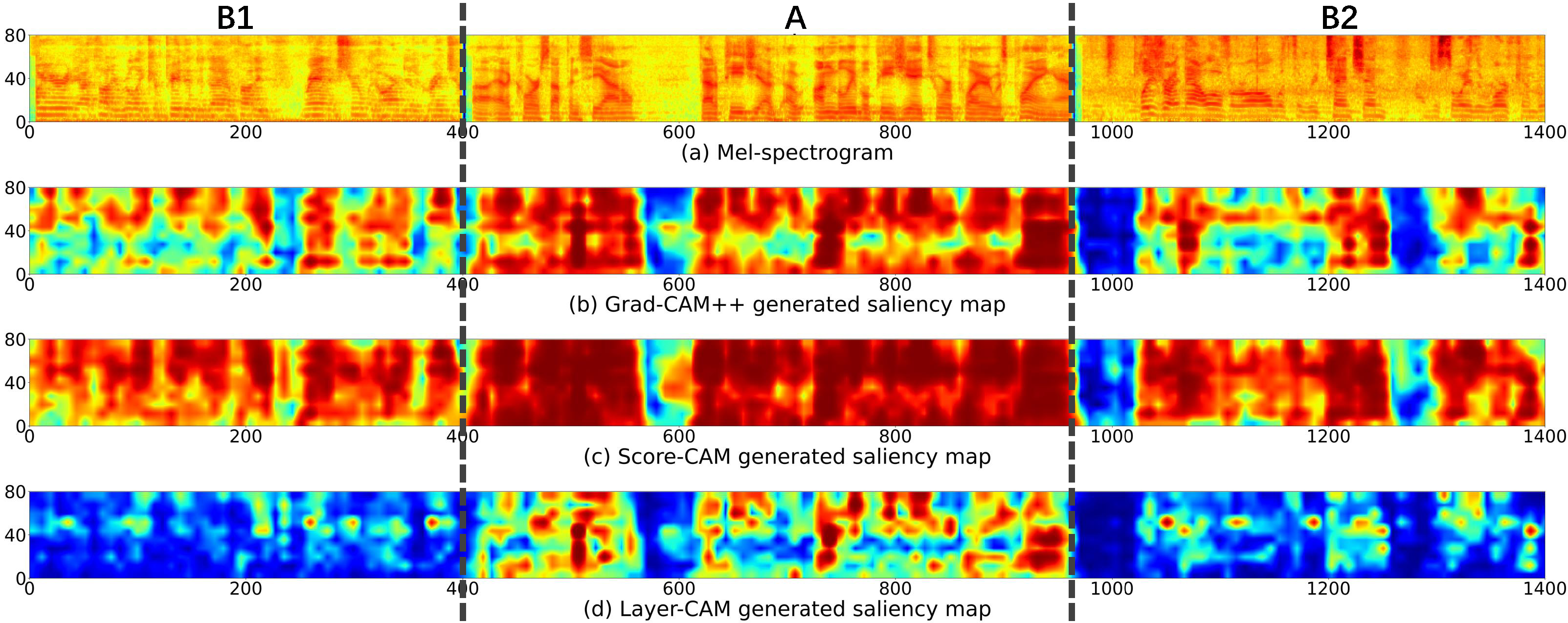}
\caption{Saliency maps on a `B-A-B' test example.}
\label{fig:test}
\end{figure}

\begin{figure}[htb!]
  \centering
  \includegraphics[width=1\linewidth]{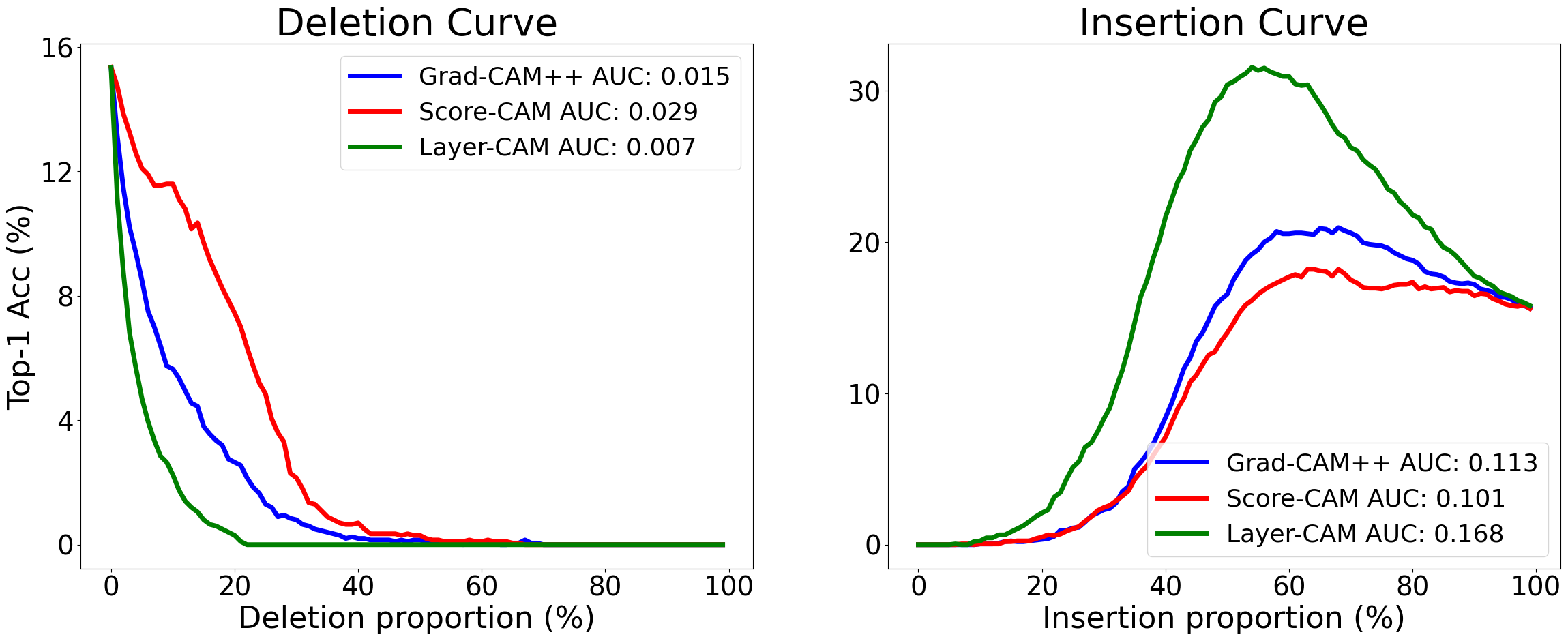}
  \caption{Deletion and insertion curves of three CAM algorithms with multi-speaker speech.}
  \label{fig:auc2}
\end{figure}

In the multi-speaker experiment, we concatenate an utterance of the target speaker with one or two utterances of other interfering speakers,
and draw the saliency map. 
Figure~\ref{fig:test} shows a `B-A-B' test example. 
A denotes the target speaker while B denotes the interfering speaker.
This time, Layer-CAM shows surprisingly good performance: it can accurately locate the 
segments of the target speaker, and mask non-target speakers almost perfectly. 
In comparison, Grad-CAM++ and Score-CAM are very weak in detecting non-target speakers. 
Moreover, Figure~\ref{fig:auc2} shows the results of the deletion and insertion curves with 2,000 multi-speaker utterances
in the B-A-B form.
It can be seen that Layer-CAM gains much better AUCs than the other two CAMs.

\subsection{Localization and recognition}


Since Layer-CAM can localize target speakers, we can use it as a tool to perform 
localization and recognition, i.e., firstly identify 
where the target speaker resides and then perform speaker recognition 
with the located segments only. We assume this is better than using the entire utterance. 

To test the hypothesis, we select 100 speakers from the training set, each with 20 utterances.
We randomly concatenate the utterances in forms A-B, B-A-B, A-B-C, and A-B-A, where A denotes target speakers
and B/C denotes interfering speakers. We use saliency maps produced at the layers of different 
ResNet blocks (S1-S4) to mask the input utterance by simple element-wised multiplication. 
Saliency maps of different layers can be combined as well. Top-1 Acc (\%) are reported in Table~\ref{tab:local}.

Firstly, we observe that neither Grad-CAM++ nor Score-CAM can achieve performance better than the 
baseline (using the whole utterance). Layer-CAM, in contrast, delivers
remarkable performance improvement, and this is the case for the saliency maps at all layers. 
This provides a very strong evidence that Layer-CAM can 
identify the important speaker-discriminative regions, while 
the other two algorithms cannot. This furthermore suggests that Layer-CAM is 
the only valid visualization tool among the three variants. 

Secondly, we find that although saliency maps at all layers produced by Layer-CAM are informative, the one from 
S2 seems the most discriminative. More investigation is required here, but one possibility is that the 
saliency map of S2 is more conservative and retains more regions when compared to the ones obtained from higher layers. 

Finally, we find that for Layer-CAM, aggregating saliency maps from different layers can improve 
performance. This observation seems consistent with the feature aggregation technique~\cite{desplanques2020ecapa,tu2022aggregating}.
Note that there is no such a trend with the other two CAMs. If we believe features are truly complimentary,
then it is another evidence that only Layer-CAM can be used as a reliable visualization tool.

\section{Conclusion}
\label{sec:con}

The ultimate goal of our study is to identify a reliable visualization tool. We focused on 
three CAM-based algorithms: Grad-CAM, Score-CAM and Layer-CAM. 
Experiments conducted with a state-of-the-art ResNet34SE model showed that 
although all the three algorithms can identify important regions in single-speaker utterances, 
Layer-CAM can localize target speakers in multi-speaker utterances. 
We therefore conclude that Layer-CAM is a reliable visualization tool for speaker recognition, 
and is the only one among the three CAM variants. In future, we will use the same protocol to test other visualization
tools. Furthermore, the localization and recognition experiments conducted here
suggests that integrating saliency maps may improve speaker recognition. 
This deserves more investigation.


\bibliographystyle{IEEEtran}
\bibliography{refs}

\end{document}